\documentclass[10pt]{iopart}
\usepackage{iopams}
\usepackage{epstopdf}
\usepackage{graphicx}
\usepackage{mathtools}
\usepackage[style=ieee, citestyle=numeric-comp, backend=biber]{biblatex}
\begin{document}

\title[Observation of a drift-driven local transport regime in the island divertor of Wendelstein 7-X]{Observation of a drift-driven local transport regime in the island divertor of Wendelstein 7-X}

\author{E Flom$^{1,}$$^2$, D M Kriete$^3$, M Krychowiak$^2$, N Maaziz$^2$, V Perseo$^2$, F Reimold$^2$, O Schmitz$^1$, V Winters$^2$, F Henke$^2$, D Gradic$^2$, R König$^2$, and the W7-X Team$^4$}

\address{$^1$ University of Wisconsin-Madison, Madison, WI, United States of America}
\address{$^2$ Max Planck Institute for Plasma Physics, Greifswald, Germany}
\address{$^3$ Auburn University, Auburn, AL, United States of America}
\address{$^4$ See author list of \cite{pedersen2022experimental}}
\ead{erik.flom@ipp.mpg.de}
\vspace{10pt}
\begin{indented}
\item[]November 2023
\end{indented}

\begin{abstract}
Measurements of the poloidal electron density and temperature distribution in the W7-X island divertor reveal the existence of two distinct scrape-off layer transport regimes. At low and intermediate densities, $T_e$ profiles across the magnetic island are hollow, with a minimum at the island O-point. Above a threshold density $\overline{n}_e \approx 5 \times 10^{19}$ m$^{-3}$, the local minimum vanishes and the $T_e$ profiles become monotonic. At the lowest studied densities, the hollow $T_e$ profiles imply an electrostatic field of up to $\textbf{E}_r \sim 6$ kV/m, which can induce $\textbf{E} \times \textbf{B}$ drifts around the island, with drift velocities of up to $\sim3$ km/s.  These $\textbf{E} \times \textbf{B}$ drifts result in strong convective heat transport poloidally around the island, binormal to the magnetic field, which supports the hollow radial temperature profile and hence provides a positive feedback to the electric field. Above the threshold density, these drifts strongly diminish, presumably due to increased anomalous diffusion, which  causes the profile hollowness and therefore also the radial electric field to strongly diminish. This argument is supported by comparison to EMC3-EIRENE modeling and an analytical transport analyses  showing that an additional, non-diffusive transport channel is required to maintain the profile hollowness up to the densities seen in experiment, and that a poloidal drift transport channel supports such hollowness only up to certain densities.
\end{abstract}

%
\vspace{2pc}

\submitto{\NF}

%
\ioptwocol

\section{Introduction}

One of the key goals of the Wendelstein 7-X (W7-X) stellarator \cite{klinger2019overview, grieger1991physics, renner2004physical} is demonstrating the feasibility of its ``island divertor" concept on the path towards a future fusion reactor \cite{pedersen2018first}. The island divertor is formed by a magnetic island chain at a low order rational surface in the plasma edge which defines the last closed flux surface. Wendelstein 7-X features ten toroidally discrete target modules, one each in the upper and lower sections of the device's five modules \cite{renner2004physical}.

The interplay between parallel and perpendicular transport in the SOL is a fundamental aspect of how well a divertor is able to serve as a heat and particle exhaust solution \cite{feng2011comparison}. Understanding the transport phenomena in the scrape-off layer (SOL) is therefore key to optimizing stellarator divertors, which are more complicated than tokamaks. This is due to the presence of much shallower magnetic field pitch angles ($\Theta$) in the SOL of W7-X than in a tokamak. The result is that in a low magnetic shear stellarator like W7-X, binormal perpendicular heat transport can be as large as the parallel transport \cite{feng2011comparison}. 

The importance of perpendicular transport processes has motivated this study into island plasma parameters. The direct measurement of local $n_e$ and $T_e$ in the island divertor domain enables the investigation of underlying transport features, which in turn allows for the investigation of overall divertor performance. In this work, care is taken to distinguish two different types of ``perpendicular'' transport, each mutually orthogonal to the local magnetic field. \textbf{\textit{(Flux-surface) perpendicular}} transport is used to denote transport along a vector which is perpendicular to the magnetic field and anti-parallel to a vector pointing towards the (island) magnetic axis. \textbf{\textit{Poloidal}} transport is used to denote transport along a vector which is perpendicular to the local field direction, but tangent to the flux surface, i.e. the binormal direction around the island. The usage of these terms is consistent with broader SOL transport literature. Making a clear distinction between these two mutually perpendicular directions is important to discuss the two different transport regimes observed and discussed in this work.

A specific focus of this work is the investigation of perpendicular transport mechanisms such as poloidal $\textbf{E} \times \textbf{B}$ drifts, which can strongly affect the heat and particle transport to the divertor targets, as they offer a shortcut between the last closed flux surface and the targets. Such drifts may affect the accessibility or stability of detached scenarios on stellarators, as has already been shown for tokamaks \cite{jaervinen2018b}. Previous work on Wendelstein 7-AS has shown up to factor of two discrepancies in particle flux onto respective upper/lower divertor targets, the asymmetry of which reversed when reversing the field direction \cite{grigull2003influence}. This behavior has been attributed to $\textbf{E} \times \textbf{B}$ drift effects \cite{feng1998modelling}. Electric fields of a strength sufficient to affect SOL transport dynamics have been shown to occur in island divertors and other edge island structures on other devices \cite{evans1989resonant, coenen2011rotation, ida2001observation}. Previously, evidence for such $\textbf{E} \times \textbf{B}$ drifts has been shown in the ``low iota'' configuration of W7-X \cite{kriete2023effects, hammond2019drift}, a study that we expand in this paper to the more common ``standard'' divertor configuration---for information about the most common magnetic configurations of the device, see e.g. \cite{sunn2017key}. 

The paper is organized as follows: in section 2, the diagnostic technique and experiments conducted are described. In section 3, the resulting measurements of the 2D $T_e(R,Z)$ and $n_e(R,Z)$ maps are presented, as well as 1D measurements in the radial direction through the island O-point. This is followed by an analysis of the radial electric field in the island domain. In section 4, the results are discussed and compared to modeling performed with the EMC3-EIRENE \cite{feng20043d}, as well as a local heat flux transport model based on the same continuity equations solved globally by the code. Finally, conclusions and ideas for future studies are  presented in section 5.

\section{Diagnostic and Description of Experiments}

\begin{figure}[h!]
    \centering
    \includegraphics[width=0.50\textwidth]{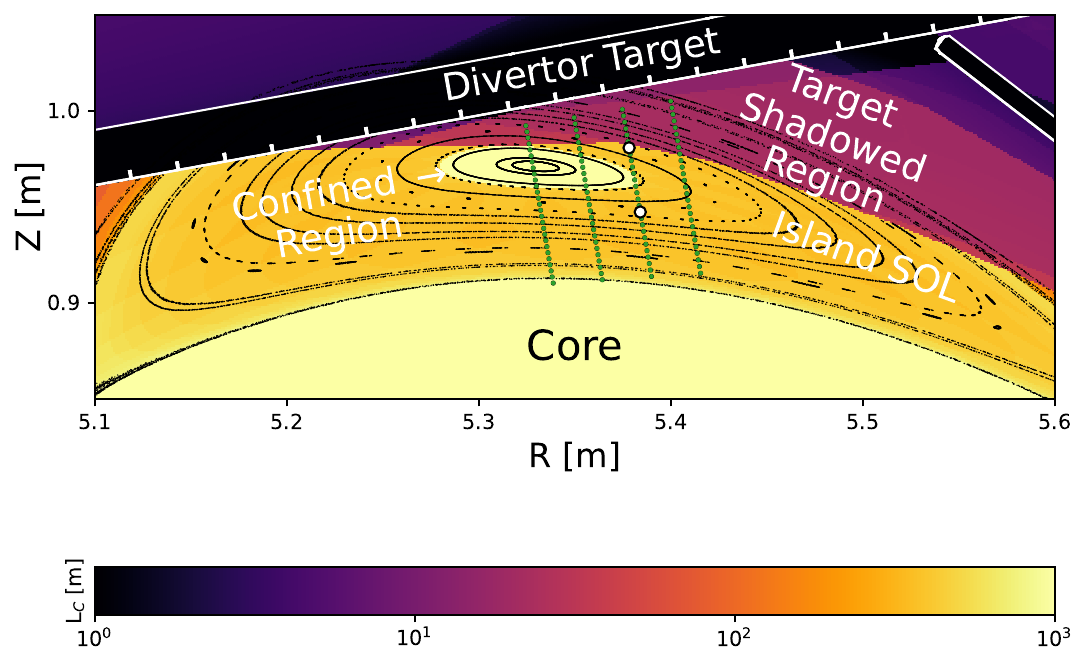}
    \caption{The viewing geometry of the divertor helium beam diagnostic on Wendelstein 7-X compared to the magnetic geometry of one of the islands in the standard magnetic configuration of W7-X, which features 5 independent islands in the edge. The green points correspond to the observation volumes of the diagnostic along the four atomic beams which were operable during these experiments. Two of the volumes are highlighted in white, corresponding to the points used for more detailed analysis presented in section 4. These two observation volumes are aligned to the same flux surface on inboard and outboard sides of the island O-point.}
    \label{fig:HeBeam Geometry}
\end{figure}

For this study, 2D radial and poloidal profiles of $T_e(R,Z)$ and $n_e(R,Z)$ were measured 
using the thermal helium beam diagnostic in the Wendelstein 7-X island divertor \cite{barbui2016feasibility, barbui2020measurements}. The measurements were taken for varying line-averaged densities, approximately $\overline{n}_e = 0.7 - 6 \times 10^{19}$ m$^{-3}$. Figure \ref{fig:HeBeam Geometry} shows the viewing geometry of the helium beam in an upper divertor of the device, the divertor of half-module 51. For further information about the island divertor geometry and the divertor targets, see e.g. \cite{renner2004physical}. In this magnetic configuration, these plasma parameter maps are available across a significant poloidal extent of the island, including the primary island SOL, the confined region around the O-point, and the target shadowed region (TSR), which features comparatively short connection lengths. Deeper discussion about the importance and influence of these different regions can be found in e.g. \cite{hammond2019drift}. These different topological regions are each annotated in the figure.

 The experiments performed consisted of a series of discharges at different line-averaged density levels and heating powers, all heated purely by electron cyclotron resonance heating (ECRH). Approximately two-second flat-top conditions at each density and power level allowed sequential activation of the four gas nozzles of the helium beam system, obtaining radial profiles at four different poloidal positions. This enabled reconstruction of interpolated $n_e(R,Z)$ and $T_e(R,Z)$ maps across this part of the magnetic island. A complete list of $\overline{n}_e$ and $P_\text{ECRH}$ values for the 17 experimental programs considered in this analysis are found in table 1.  $\overline{n}_e$ was determined by a dispersion interferometer providing a line-integrated measurement through the plasma core. This was then divided by the 1.33 m path length through the plasma to get a line-averaged measurement, consistent with other works \cite{brunner2018real}. 
 
All of the plasmas analyzed in this work are attached to the divertor target, as characterized by a radiative fraction $f_\text{rad}$ of below $50 \%$, a typical threshold at which the first signs of detachment begin in W7-X plasmas. This threshold has been validated both experimentally \cite{perseo20212d, schmitz2020stable} and with EMC3-EIRENE modeling \cite{feng2021understanding}. As such, it is not expected that the radiation dynamics within the island are significantly different between the different plasmas analyzed in this work, with most radiation occurring at the target very near the strike line position, and only small radiative losses in the rest of the island. Therefore, in this work the primary differences in the profiles are assumed to be due to the differing energy sourcing behavior into the island bulk, i.e. the evolving ratio of perpendicular and parallel transport, rather than strongly varying energy sink terms.

\hfill
\begin{table}[h!]
\centering
\label{table:discharges}
\begin{tabular}{|c | c | c |} 
 \hline
   $\overline{n}_e$ [$10^{19}$ m$^{-3}$]& $\text{P}_\text{ECRH}$ [MW] & Program nos.\\ 
 \hline\hline
 0.7 & 1.0 & 51-52 \\ 
 \hline
 1.1 & 1.5 & 53-55 \\
 \hline
 1.5 & 2.0 & 56, 58\\ 
 \hline
 2.3 & 2.5 & 59-60\\
 \hline
 3.0 & 3.0 & 63-66\\
  \hline
 5.0 &  3.5 & 67-68\\
   \hline
 6.0 & 4.0 & 69-70\\
\hline
\end{tabular}
\caption{A table of the 17 W7-X plasmas considered in this study. Density feedback was to within $\Delta \overline{n}_e = 0.2 \times 10^{19}$ m$^{-3}$ and $P_{ECRH}$ was maintained to within $250$ kW. W7-X Program numbers are given to fulfill the format 20230117.xx}
\end{table}

\section{Experimental Observations}

\subsection{Measurements of 2D Temperature Profiles}

\begin{figure*}
    \centering
    \includegraphics[width=0.48\textwidth]{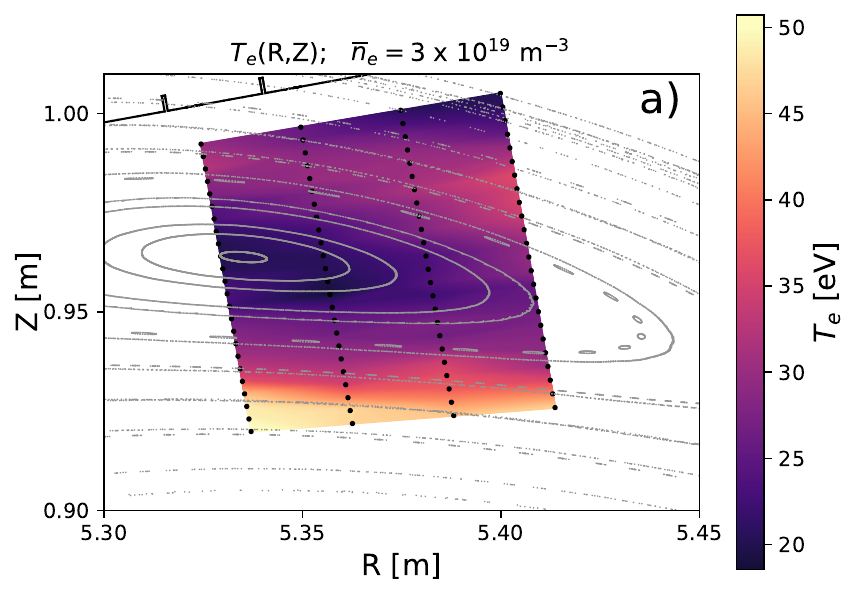}
    \includegraphics[width=0.48\textwidth]{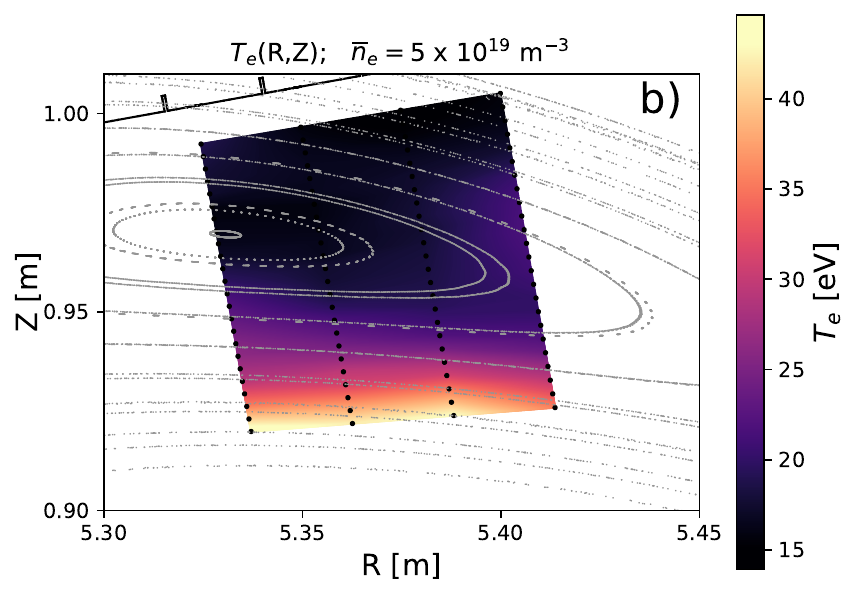}
    \includegraphics[width=0.48\textwidth]{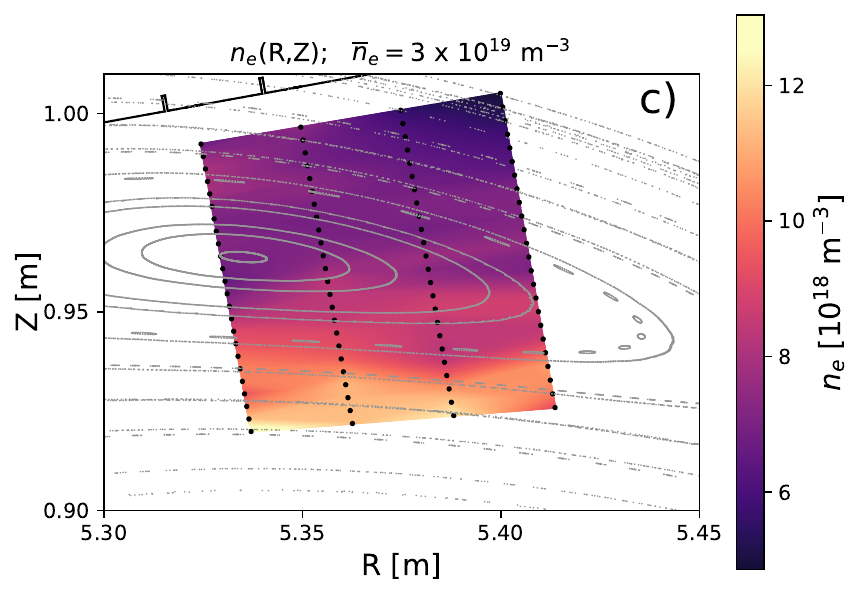}
    \includegraphics[width=0.48\textwidth]{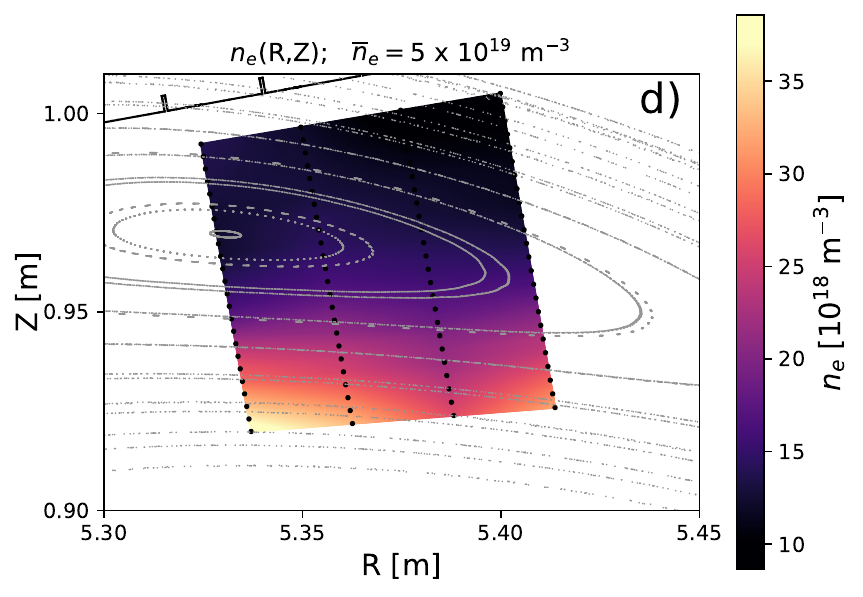}
    \caption{Electron temperature (upper row) and density (lower row) maps as measured by the helium beam diagnostic in the upper divertor of Wendelstein 7-X for experimental conditions of $\overline{n}_e = 3 \times 10^{19}$ m$^{-3}$ (left column) and  $\overline{n}_e = 5 \times 10^{19}$ m$^{-3}$ (right column). The visible differences in the Poincare plots between the two densities comes from the additional toroidal boostrap current in the lower density experiments, which has been included in the magnetic reconstruction via a filament-on-axis model. Note the different color bar scaling for each image.}
    \label{fig:HeBeam_12e18}
\end{figure*}

In figure \ref{fig:HeBeam_12e18}, the electron temperature and density maps $T_e(R,Z)$ and $n_e(R,Z)$ across the magnetic island at one toroidal angle are shown for two density levels. $T_e(R,Z)$ features for the low density level at $\overline{n}_e = 3 \times 10^{19}$ m$^{-3}$ a temperature hole with a temperature minimum located at the center of the island O-point (figure \ref{fig:HeBeam_12e18}a), while this feature vanishes to the level of the diagnostic noise at the higher density of $\overline{n}_e = 5 \times 10^{19}$ m$^{-3}$ (figure \ref{fig:HeBeam_12e18}b). In the $\overline{n}_e = 3 \times 10^{19}$ m$^{-3}$ case, the electron temperature profile within the SOL shows clear conformity to the magnetic contours of the island, while the higher density case of $\overline{n}_e = 5 \times 10^{19}$ m$^{-3}$ shows electron temperature profiles where the hollowness has fallen to be on the level of the diagnostic uncertainty. In both examples, the density profiles fall off radially and do not exhibit a characteristic pinning to the island flux surfaces like the temperature profiles do at low- and moderate densities (figure \ref{fig:HeBeam_12e18}c). However, the overall edge density gradients are higher at higher density (figure \ref{fig:HeBeam_12e18}d), which as discussed later in the paper is key to driving increased anomalous diffusive transport. 

\begin{figure}[h!]
    \centering
    \includegraphics[width=0.5\textwidth]{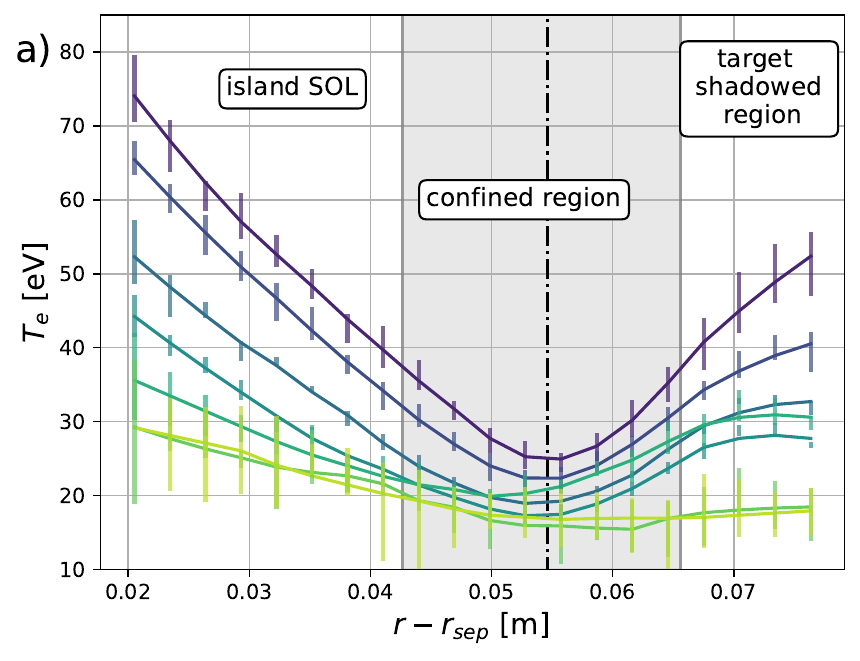}
    \includegraphics[width=0.5\textwidth]{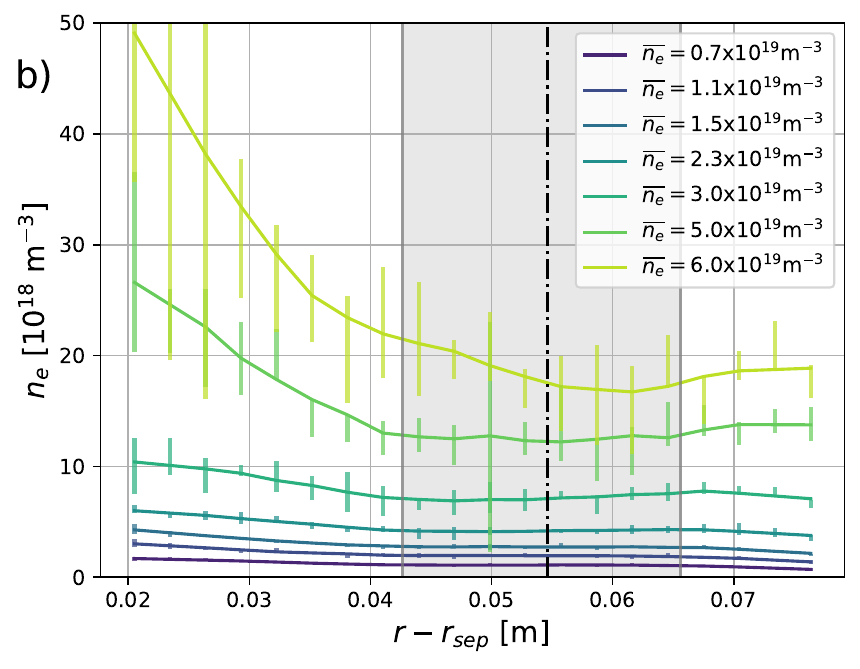}
    \caption{1D electron temperature profiles (above) and electron density profiles (below) through the island O-point for seven points of a density scan performed on Wendelstein 7-X. The color coding of the traces is the same as presented in figure \ref{fig:e-field}.}
    \label{fig:1D_Te}
\end{figure}

 To illustrate the evolution of profile hollowness, one-dimensional $T_e$ profiles acquired radially through the island O-point are presented in figure \ref{fig:1D_Te}. The general trend observable in this figure is that edge temperatures, including separatrix temperatures, decrease systematically with increasing $\overline{n}_e$, while the temperature minimum for hollow profiles always remains consistent with the location of the island O-point. The error bars in this figure are comprised of the fitting uncertainty in the spectroscopic analysis of the measured line radiation and statistical errors from repeated measurements at nominally the same heating power and density. Previous work has shown that large ($>$30$\%$) systematic error bars exist due to uncertainties on the underlying atomic data \cite{flom2022bayesian}, but these systematic errors are neglected here as they would predominantly result in consistent shifts of the $T_e$ and $n_e$ profiles with minimal effect on profile shape, and they do not affect the relative uncertainty. A small inward shift is visible in the $\overline{n}_e = 3 \times 10^{19}$ m$^{-3}$ trace, and is attributed to a larger than usual toroidal (bootstrap) current compared to other discharges considered in this study, causing an inward shift of the islands \cite{gao2019effects}. This temperature hollowness persists for all densities up to $\overline{n}_e = 3 \times 10^{19}$ m$^{-3}$, until there is a significant drop in the temperature radially outwards of the O-point, leading to monotonic temperature profiles.  

\subsection{Estimates for an island-radial electric field derived from the measured hollow $T_e$ profiles}
The hollow temperature profile suggests an electric field $\textbf{E}_r$ which points towards the island O-point at all points within the island. To arrive at this conclusion, we assume that $\Phi_{plasma} \approx 3T_e/e$ at all points along a field line far from the target \cite{stangeby1996simple}. This implicitly assumes that the temperature along the field line is constant, i.e. $T_e \approx T_{target}$, and uses a sheath boundary condition to set the field line potential. In the presence of large temperature gradients along the field line, the actual local potential values derived from this model may be inaccurate. This model also requires that no large current densities are present in the SOL. These are moderately strong assumptions in the SOL of Wendelstein 7-X, but for an order-of-magnitude estimate likely suffice, as will be discussed subsequently. Under these assumptions, the radial electric $E_r$ can then be calculated from the radial gradient of the electron temperature, and the resulting field points towards the island O-point from all points in the island in the case of hollow $T_e$ profiles. Care must be taken here to not confuse machine radial and island radial coordinates. In the geometry of W7-X, the electric fields inferred in this work point perpendicular to the \textit{island} flux surfaces, which are not always exactly perpendicular to the \textit{core} flux surfaces.

\begin{figure}[h!]
    \centering
    \includegraphics[width=0.5\textwidth]{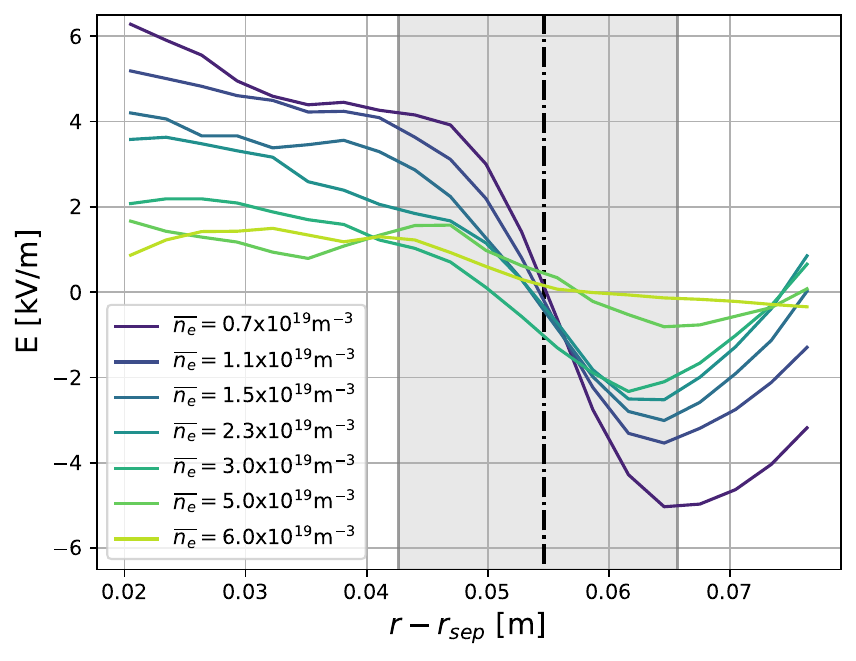}
    \caption{$\textbf{E}_r$ profiles derived using the simple $\Phi = 3T_e/e$ model for seven points of a density scan performed on Wendelstein 7-X, colors corresponding to the $T_e$ profiles presented in fig. \ref{fig:1D_Te}. The color coding of the island SOL, confined region, and TSR are retained as well.}
    \label{fig:e-field}
\end{figure}

Using this $\Phi \approx 3T_e/e$ Ansatz, radial electric field strengths were computed for several discharges of increasing $\overline{n}_e$, shown in figure \ref{fig:e-field}. These $\textbf{E}_r$ profiles are relatively flat towards the separatrix and decrease in magnitude with increasing $\overline{n}_e$ from approximately 6 kV/m to 1 kV/m. Into the confined region, steep electric field profiles are inferred, with a zero crossing consistently at the island O-point. This behavior holds for all density cases up to $n_e \approx 5.0 \times 10^{19}$ m$^{-3}$. Above this density, characteristic hollow profile shape outside of the magnetic O-point (indicated as a dot-dashed line) is lost and the profile shifts suddenly to much smaller values overall. This is consistent with other experimental observations and will be compared in section 4 to numerical and analytical analyses of the SOL heat transport with and without drift effects. 

Profile hollowness is driven by the balance of sources and sinks between the different flux surfaces. Sources include perpendicular diffusion through the island vs. parallel/poloidal convection around the island, while sinks include i.e. local ionization and radiation balances. It is obvious that cross field transport has to provide the energy source for ionisation of the main species, creating the island plasma in the first place. This means that a balance of perpendicular heat influx and local energy losses is established early on that may set the temperature profiles. If this induces profile hollowness, the setup of the electric field we described before can have a self-reinforcing effect. Even in the case where no profile hollowness exists to begin with, strong perpendicular gradients may induce strong drifts which distort the island $T_e$ profile shapes, again leading to a stable hollow profile shape. The $\textbf{E} \times \textbf{B}$ drift-induced transport channels heat around the island and reduces the available perpendicular energy source to the island center. This contributes to or may even enhance the hollow temperature profile which, in turn, supports the radial electric fields. While the energy sinks within the island are finite, there has been no observed accumulation of strong radiative features at the island O-point during attached plasmas in this magnetic configuration, either from tomograms of carbon emission in the divertor islands \cite{krychowiak2021gaussian} or from EMC3-EIRENE modelling \cite{schmitz2020stable, feng2021understanding}. As such, the source of changing profile hollowness in these experiments is due to a changing relationship between the perpendicular and poloidal heat transport channels in the SOL and not strongly changing local energy sinks.

Flux-surface perpendicular diffusive transport becomes increasingly important at higher SOL densities and reduced SOL temperatures, as will be shown in section 4 and the references contained therein.  In summary, $q_\perp \propto \chi_\perp n + D_\perp  \nabla n$, so for fixed transport coefficients, as is assumed in this work, the net perpendicular heat flux increases with density. This is due to the rise of strong perpendicular density gradients and which enable efficient diffusive transport into the island O-point flux bundles, and is consistent with the increasing edge density gradients shown in figure \ref{fig:HeBeam_12e18}. Between the two experimental densities shown in figure \ref{fig:HeBeam_12e18}, it is hypothesized that a sufficiently shallow temperature gradient is reached between the flux surfaces of the island. This coincides with the perpendicular density gradient increasing, seemingly pushing the ratio $q_\perp / q_\parallel$ past some threshold value, at which point the self-reinforcing behavior of the drifts breaks down and the temperature profiles in the island transition to be radially-decreasing rather than hollow.

The electric field model applied here applies a sheath boundary condition for the plasma potential. This is not typically applicable along the entire flux tube, and is therefore not intended as a means to derive precise values of $\textbf{E}_r$. However, there is a general agreement between this model and previously published SOL measurements from other diagnostics, modulo a small factor. For example, inferred $\textbf{E}_r$ in this work on the order of several kV/m imply poloidal $\textbf{E} \times \textbf{B}$ drift velocities on the order of $v_{\textbf{E} \times \textbf{B}} = \textbf{E} \times \textbf{B} / B^2 \approx$ $1$-$3$ km/s poloidally around the island. Previous works have arrived at similar $v_{pol}$ values of $2$-$5$ km/s near the separatrix \cite{killer2020plasma} and similar $\textbf{E}_r$ values of $5$-$15$ kV/m at the separatrix  by independent measurement and the assumption that $v_{\text{pol}} = v_{\textbf{E} \times \textbf{B}}$ \cite{carralero2020characterization}. Also, in \cite{alonso2022plasma} a transition from ion root (negative $\textbf{E}_r$) just inside the LCFS to a large, positive $\textbf{E}_r$ just outside the LCFS was presented.  All of these measurements are consistent in magnitude and direction with the ones presented in this work. This supports that the sheath boundary model is sufficient to inform particle fluxes and associated convective heat fluxes, at worst on an order-of-magnitude level, likely to within a smaller factor. The order-of-magnitude arguments presented in section 4.2 are then not significantly changed by the the simplifications induced by this sheath potential model, and the overall conclusion remains: that the $\textbf{E} \times \textbf{B}$ drift effects are a significant part of the overall transport dynamics in the standard magnetic configuration of W7-X.

\section{Discussion}

We have shown that the inferred electric fields and associated drift velocities are consistent with other measurements in various regions of the SOL. The question still remains, however, if these drifts are capable of producing or at least reinforcing profile hollowness. In this section, we investigate the feasibility of two key principles: 1) that the observed hollow $T_e$ profiles are plausibly drift-supported and 2) that, at some point, perpendicular diffusive transport caused by increasing edge density gradients is plausibly leading to a loss of $T_e$ profile hollowness, which ends the positive feedback cycle of poloidal drifts maintaining deeply hollow profiles. These two points are investigated via a combination of EMC3-EIRENE modeling and analytical heat flux estimations for local transport features. 

\subsection{Numerical analysis of required diffusive transport levels in the SOL to maintain hollow $T_e$ profiles using EMC3-EIRENE}

\begin{figure}[h!]
\centering
\includegraphics[width=0.50\textwidth]{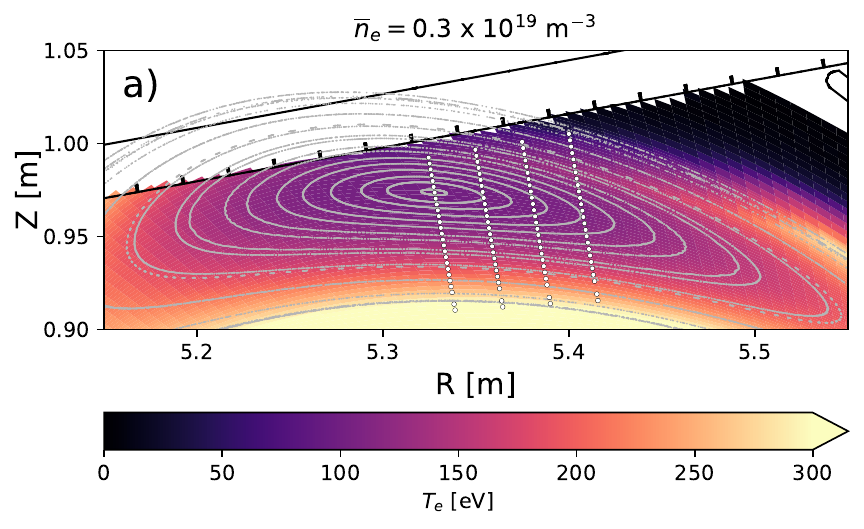}
\includegraphics[width=0.50\textwidth]{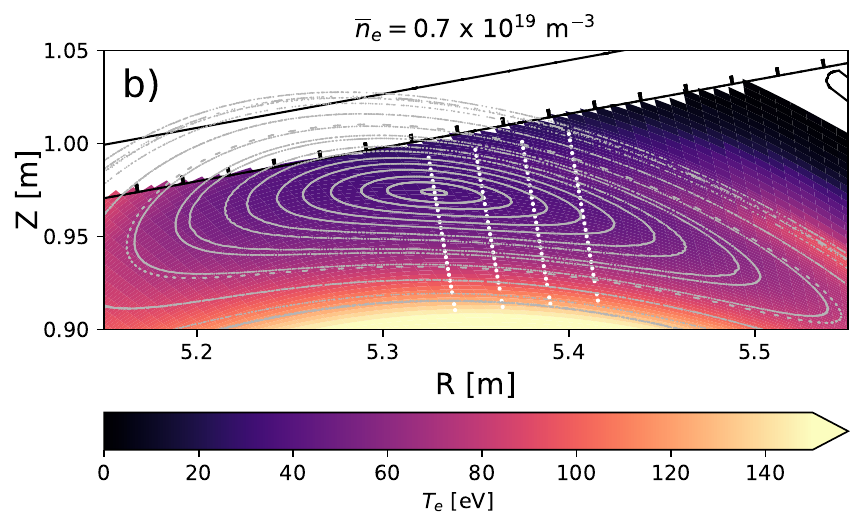}
\includegraphics[width=0.50\textwidth]{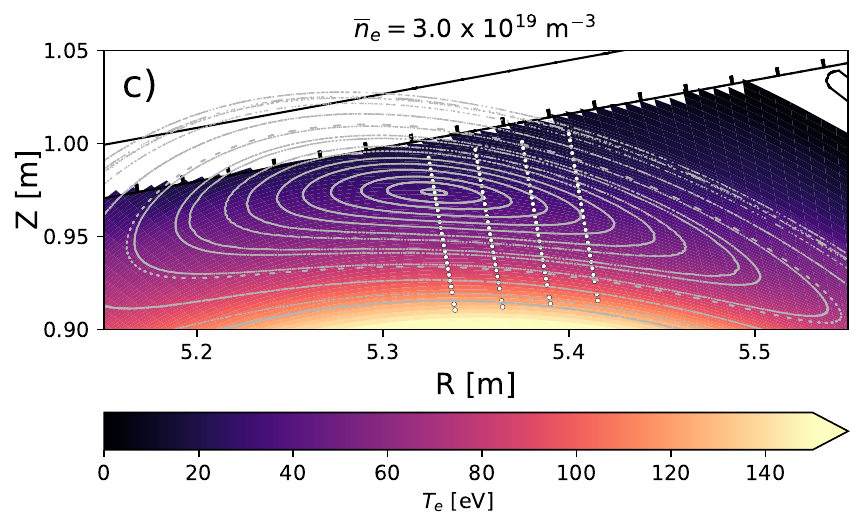}
\caption{Electron temperatures as modeled by EMC3-EIRENE in the upper divertor of Wendelstein 7-X for $n_{sep}$ = $1 \times 10 ^{18}$ m$^{-3} \approx \overline{n}_e = 0.3 \times 10^{19}$m$^{-3}$ and $P_{ECRH}$ = 1 MW (top); $n_{sep}$ = $0.24 \times 10 ^{18}$  m$^{-3} \approx \overline{n}_e = 0.7 \times 10^{19}$m$^{-3}$ and $P_{ECRH}$ = 1 MW (middle); and $n_{sep}$ = $9 \times 10 ^{18}$  m$^{-3} \approx \overline{n}_e = 3 \times 10^{19}$ m$^{-3}$ and $P_{ECRH}$ = 3 MW (bottom). For clarity, the colormap for the top image has been scaled by a factor of 2.}
\label{fig:EMC3_1e18}\end{figure}

The EMC3-EIRENE code solves the 3D Braginskii fluid equations in a diffusive transport Ansatz without drift effects \cite{feng20043d}. We use this standard tool for stellarator boundary modeling here to predict the 2D $T_e(R,Z)$ maps measured and assess the level of diffusive transport required to reconstruct the measured profiles. Synthetic 2D $T_e$ maps in the same region as shown in figure \ref{fig:HeBeam_12e18} are qualitatively analyzed. These comparisons can provide insight into the role of drifts in the plasma edge transport and modeling.  For this work, the anomalous transport coefficients $D_\perp$ and $\chi_\perp$ were adjusted to match experimentally measured temperature- and density fall-off lengths $\lambda_{T_e}$ and $\lambda_{n_e}$ in the region between the last closed flux surface and the island O-point. The resulting best fit was found for isotropic values of $D_\perp = 0.2$ m$^2/$s and $\chi_\perp = 1.5$ m$^2/$s. These values were used for all EMC3-EIRENE simulations in this work. Previous measurements with Langmuir probes in the SOL have resulted in similar values of $D_\perp$ in the range 0.1-0.2 m$^2$/s \cite{killer2021turbulent}. The fitted ratio of $\chi_\perp = 7.5D_\perp$ departs somewhat from the typically used $\chi_\perp = 3D_\perp$, but other works have shown the difficulty in simultaneously matching both separatrix and target parameters under this assumption \cite{bold2022parametrisation}. For all simulations, radiated power was fixed to the level observed in experimental density scan, approximately $10$-$25 \%$ of the total heating power. 

First, an attempt was made to reproduce the hollow temperature profiles seen in experiment by matching experimental conditions directly. The resulting set of EMC3-EIRENE simulations as $T_e(R,Z)$ distributions is shown in figure \ref{fig:EMC3_1e18}, with increasing line-averaged density from top to bottom of the figure. The density values chosen for this modeling are close to the density values from experiment, as shown in the lower two panels of figure \ref{fig:EMC3_1e18}. The density input used for the results shown in the central panel correspond approximately to the lowest experimental density, $n_{sep} = 0.24 \times 10^{19}$ m$^{-3} \approx \overline{n}_e = 0.7 \times 10^{19}$ m$^{-3}$. Here, the $T_e$ profiles in the island are only very slightly hollow, with the temperature contours only beginning to follow the magnetic surfaces near the island X-points, and not significantly into the observation region of the diagnostic. In the bottom panel of figure \ref{fig:EMC3_1e18}, at higher density, the temperature contours are nearly independent of the magnetic flux surfaces of the island and instead show primarily a radially decay outward from the last closed flux surface. This simulation was performed for $n_{sep} = 1.0 \times 10^{19}$ m$^{-3} \approx \overline{n}_e = 3.0 \times 10^{19}$ m$^{-3}$, and shows significant difference in profile behavior to experimental measurements at similar densities as shown in figure \ref{fig:HeBeam_12e18}, in which the measured profiles are still clearly hollow.

As seen in panels b and c of this figure, the hollow temperature profiles observed could not be reproduced with density and heating parameters matched directly to experiment.  Instead, it was necessary to reduce $\overline{n}_e$ approximately a factor of two below the experimentally achievable value, for the same $1$ MW heating power as applied in experiment. For EMC3-EIRENE simulations to show hollowness on the level seen in experiment, it was necessary to stay approximately a factor of ten below the experimental density presented in the figure \ref{fig:HeBeam_12e18}a, where the temperature profile is still clearly hollow. A simulation was conducted at very low density ($n_{sep}$ = $1 \times 10 ^{18}$ m$^{-3} \approx \overline{n}_e = 0.3 \times 10^{19}$ m$^{-3}$) as shown in figure \ref{fig:EMC3_1e18}a. Here, the island isotherms are seen to curve around the island in a similar manner to what has been observed in experiment. Note that due to the high heating power and low density required to achieve this level of profile hollowness, the modeled separatrix temperature is much higher than is seen in experiment, providing further indication that EMC3-EIRENE is not fully reflecting the perpendicular transport in the SOL.

This observation can, within the EMC3-EIRENE framework, be interpreted as follows: in the presence of a strong poloidal convective channel, plasma would be able to travel around the island along a much shorter distance compared to the parallel field line length in the island flux tube that connects to divertor targets. The ratio between the poloidal and parallel distances is a field-line averaged $\Theta$. Due to the small $\Theta$ present in the islands of W7-X (which can take values between $10^{-2}$ and $10^{-3}$, depending on configuration and location within the island), even a relatively modest poloidal velocity may be able to compete with the parallel convection and parallel conduction. This poloidal convective channel due to $\textbf{E} \times \textbf{B}$ drifts is not considered in EMC3-EIRENE. As such, hollow profiles can only be generated by increasing the parallel transport relative to the radial diffusive transport. This allows the helicity of the field lines to bring high temperature plasma around to the other side of the magnetic island over hundreds of meters of field line length. Assuming fixed transport coefficients, modifying $q_\parallel / q_\perp$ is then possible by e.g. pushing the simulation to very low $\overline{n}_e$ values and very high separatrix temperatures. This is emphasized by the fact that we rely on the $T^{5/2}$ scaling of parallel conductivity to transfer large amounts of heat in the parallel direction \cite{stangeby2000plasma}. 

This analysis shows that strongly hollow temperature profiles across the divertor island require relatively small radial transport compared to parallel transport. Due to the experimental observation of vanishing profile hollowness at high line-averaged densities, we now explore the impact of higher edge densities in EMC3-EIRENE simulations. It should be noted that hollow $T_e$ profiles could be achieved in EMC3-EIRENE at moderately higher densities with the application of strongly-varying transport parameters within the island, namely the suppression of $\chi_\perp$ in the closed field-line region of the island by a factor of 100. However, such suppression also resulted in poorer fits at higher density and the physical motivation for such a transport profile variation is unclear. As a result, we have retained isotropic transport coefficients for the entirety of this work.

The exact balance of parallel, perpendicular, and radial transport channels required to preserve a specific level of hollowness in the $T_e$ profiles is difficult to reconstruct given the current capabilities of EMC3-EIRENE. However, the requirement to go to very low density values in order to match the measured $T_e$ profile hollowness seen in experiment indicates that a poloidal transport channel must exist which is not considered in EMC3-EIRENE. We propose that this channel is comprised of the poloidal $\textbf{E} \times \textbf{B}$ drifts that we inferred from the measured profile hollowness. To further evaluate the plausibility of this argument, five characteristic heat transport terms are calculated and presented in the next section.  

\subsection{Quantitative estimates of $q_\perp$ and $q_\parallel$ scaling from experimentally measured and simulated profiles}

In this section we compare local heat fluxes calculated for five different heat transport channels: parallel convection, parallel conduction, anomalous radial diffusion, anomalous poloidal diffusion and poloidal $\textbf{E} \times \textbf{B}$  drift transport, abbreviated $q_{\parallel, \text{conv.}}$, $q_{\parallel, \text{cond.}}$, $q_{\perp \text{, anom.}}$, $q_{\text{pol., anom.}}$, $q_{\text{pol., drift}}$, respectively. The parameterizations used in this work for each are:

\begin{equation}
q_{\parallel, \text{conv.}} = \frac{5}{2} n_eT_ec_sM
\end{equation}
\begin{equation}
q_{\parallel, \text{cond.}} = -\kappa_0 T_e^\frac{5}{2}\nabla_\parallel T_e 
\end{equation}
\begin{equation}
q_{\perp \text{, anom.}} = -n_e \chi_\perp  \nabla_\text{rad.} T_e -\frac{5}{2}T_e D_\perp \nabla_\text{rad.} n_e 
\end{equation}
\begin{equation}
q_{\text{pol.}, \text{anom.}} = -\Theta^{-1} (n_e \chi_\perp \nabla_\parallel T_e + \frac{5}{2} T_e D_\perp \nabla_\parallel n_e) 
\end{equation}
\begin{equation}
q_{\text{pol., drift}} = \frac{5}{2} T_en_ev_{E \times B}
\end{equation}

\begin{figure}[h!]
    \centering
    \includegraphics[width=0.5\textwidth]{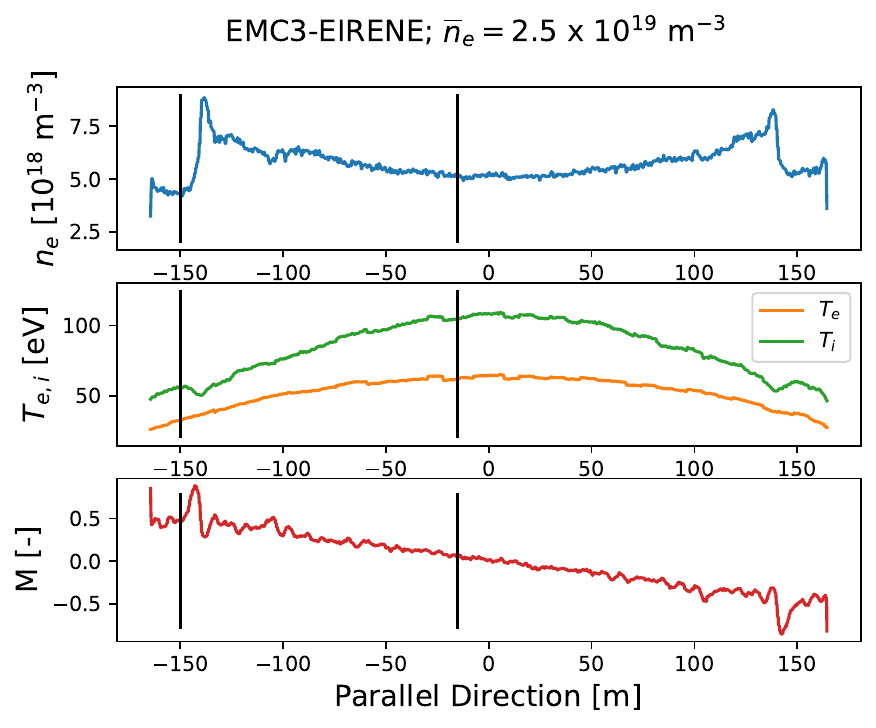}
    \caption{Plasma parameters along the field line which connects the two white points in figure \ref{fig:HeBeam Geometry}. The approximate positions of these two helium beam observation points along the field line are marked in black lines.}
    \label{fig:emc3_parallel}
\end{figure}

 The first four of these equations are simplified versions of the same heat transport equations solved by EMC3-EIRENE, while the final equation \cite{stangeby2000plasma} is not considered in the current EMC3-EIRENE model. For this comparison, the local values of $n_e$, $T_e$, $\nabla_\parallel T_e$, $\nabla_\perp n_e$, and $v_{\textbf{E} \times \textbf{B}}$ are taken from the radially outward-most of the two white reference points shown in figure \ref{fig:HeBeam Geometry}. The ion sound speed, $c_s$, is calculated using the assumption $T_i = 2T_e$ \cite{huber2000spectroscopic}, and the Mach number $M$ is approximated as $0.3$. Parallel gradients were estimated by measuring $T_e$ and $n_e$ at two locations on the same flux surface, approximately halfway between separatrix and island O-point, on two opposite sides of the island (see the description of figure \ref{fig:HeBeam Geometry} and the highlighted observation volumes therein). The difference in parameters was then divided by the field line length between the two locations to get a first-order approximation of the gradients. In reality, the field line temperature is likely not monotonic and most likely not linear. In order to support these assumptions, EMC3-EIRENE simulations of $n_e$, $T_e$, $T_i$, and Mach number profiles along the magnetic field line are presented in figure \ref{fig:emc3_parallel}. These profiles show that the approximation of parallel gradients as linear, as well as the choice of Mach number and ratio $T_i / T_e$ is well-supported by the modeling results. Conversely, the parallel density gradient is found to be non-linear but features a profile shape determined by downstream sources, similar to expectations from Stangeby's simple SOL model \cite{stangeby2000plasma}. This likely introduces a significant inaccuracy in the estimation of the heat fluxes for the poloidal convective term. In particular, the density gradient pointing downstream would then lead the density gradient term of eq. 4 to point towards the upstream point, while the temperature gradient term would point towards the downstream point. Therefore, the estimations from eq. 4 are to be taken as an upper limit. We estimate from the above argument the uncertainty in the heat flux terms to be independent from each other and within a factor of two for each term. Therefore, we discuss in our conclusion the overall trends of each heat flux term and not the absolute magnitude. 

\begin{figure}[h!]
    \centering
    \includegraphics[width=0.50\textwidth]{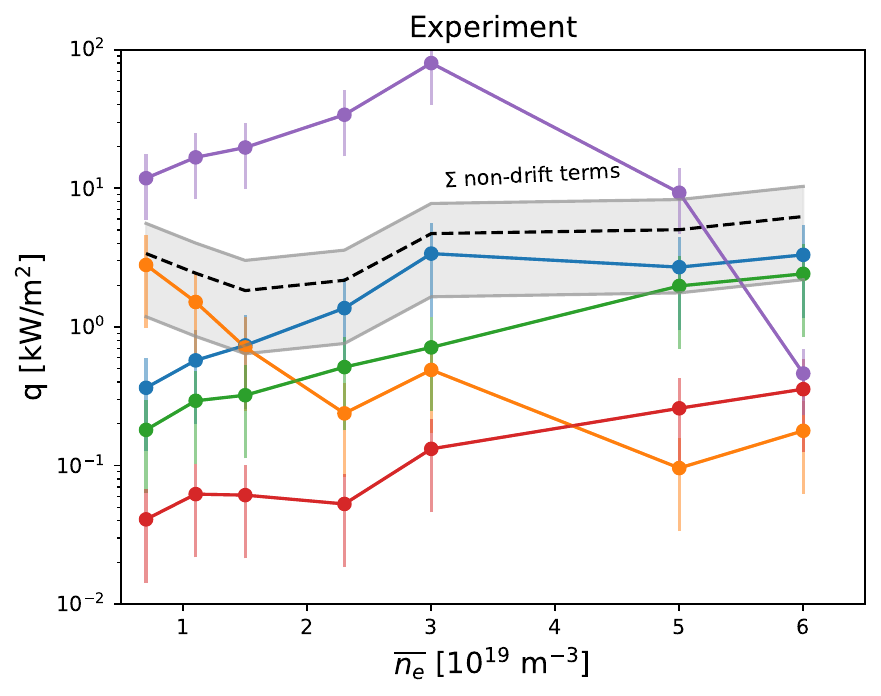}
    \includegraphics[width=0.50\textwidth]{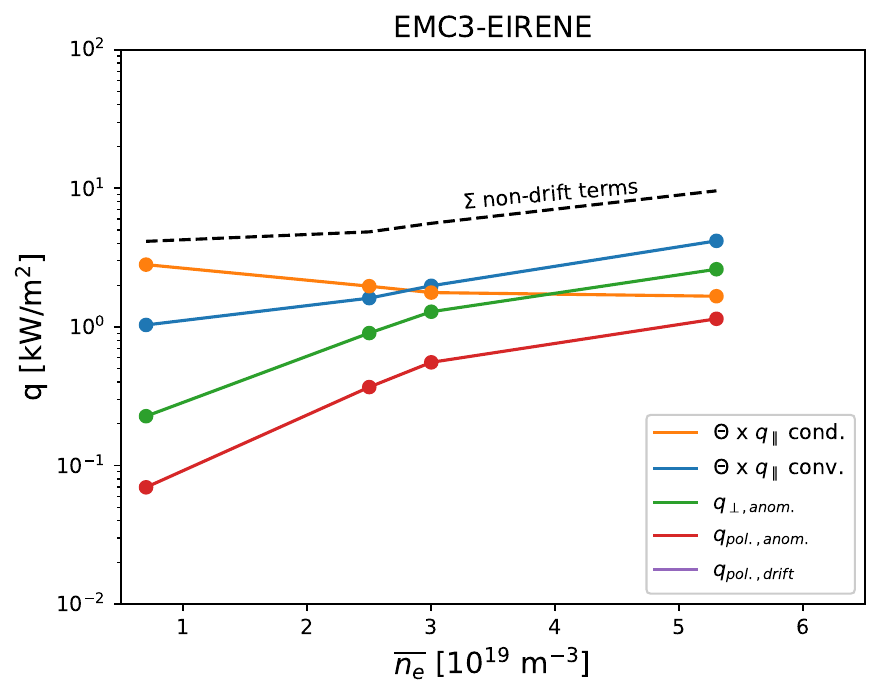}
    \caption{Estimates for the five heat transport terms outlined in Eqs. 2-6 for experimental profiles (above) and profiles generated with EMC3-EIRENE (below). The sum of the non-drift terms is shown with a dashed black line.}
    \label{fig:transport_estimate}
\end{figure}

 For each point in the density scan presented in table 1, the five heat fluxes in equations $1$-$5$ are calculated and plotted at the location of the outwardmost of the two volumes highlighted in figure \ref{fig:HeBeam Geometry}. The results of these calculations are plotted in figure \ref{fig:transport_estimate}. Transport coefficients used in these estimations are $D_\perp = 0.2$ m$^2$/s and $\chi_\perp = 1.5$ m$^2$/s, consistent with the EMC3-EIRENE simulations presented in section 4.1. The two parallel heat fluxes are additionally scaled by $\Theta = 10^{-3}$ to map between the different path lengths parallel and poloidally around the islands, as previously discussed \cite{feng2011comparison}.  Error bars are generated by linearly propagating an estimated $30 \%$ error for both $n_e$, $T_e$, as well as their gradients, consistent with latest modeling work which has estimated the impacts of uncertain atomic data on the resulting $T_e$ and $n_e$ calculations from helium beam diagnostics \cite{flom2022bayesian}. The uncertainty on the $\textbf{E}_r$ and therefore the $q_{\text{drift}}$ term is taken to be $50 \%$, due to both an underlying uncertainty in the radial temperature gradients ($30 \%$, by the same argument as above) and the small temperature anisotropies along the field line neglected in this simple model. 
 
From these calculated heat fluxes, a clear trend is seen: at low density, the dominant heat transport mechanism is from the drift term. This term is reduced  by approximately an order of magnitude when approaching higher line-averaged density. Simultaneously, the perpendicular anomalous heat convection is increased due to its linear scaling in density by a similar factor, changing the ratio $q_{\perp \text{, anom.}}/q_{\text{drift}}$ by a factor of $\sim$100 as $\overline{n}_e$ is increased by a factor of 8. The results of this analysis, within the selections we made for the experimental data used, support the previous explanation for the hollow-to-monotonic profile transition. At low density, the $\textbf{E} \times \textbf{B}$ drift transport represents the dominant perpendicular heat flux term in the island SOL, preserving the hollow temperature profiles by the method previously described. At intermediate densities, perpendicular diffusion increases the influx of energy into the island O-point and reduces the $T_e$ profile hollowness. This causes weaker radial electric fields, which in turn result in reduced poloidal convection. Therefore, at higher densities, the profile hollowness vanishes. 

It is important to note that the two parallel transport terms depicted in figure \ref{fig:transport_estimate} scale versus $\overline{n}_e$ in a way which is perhaps not immediately obvious from eq. 1 and 2. While the convective term increases with density, as would be expected from eq. 1, the conductive term decreases in magnitude with density, when no specific density dependence is apparent from eq. 2. This is a result of the specific experimental conditions for these discharges, namely heating power availability cap of approximately 4.5 MW. This led the density $\overline{n}_e$ to be increased faster than the heating power, leading to a progressive drop in SOL temperatures, as shown in figure \ref{fig:HeBeam_12e18}a. Because of this inverse experimental relationship between the upstream temperature and the line-averaged density, caused by the available heating power for this scan, the conductive heat flux along the field lines is reduced with increasing density. These two trends result in a progressive increase in the ratio of convective to conductive heat flux with increasing plasma density. In a situation where heating power per particle were kept constant, an increasing importance of conductive parallel transport may be expected with increasing SOL density in order to supply the downstream ionization source with power as the parallel convective channel saturates, e.g. the high recycling transport regime of tokamak discharges \cite{stangeby2000plasma}. Alternatively, some modeling results have shown that due to strong anomalous transport at high density, separatrix temperatures in W7-X are prevented from climbing high enough to enable a conduction-dominated regime \cite{feng2021understanding}. 

In order to compare to the trends in the heat flux features around the island from experiment, equations 1-4 were also evaluated on the results of a four point density scan from EMC3-EIRENE. The density values and heating powers were comparable to the experimental conditions. The result is shown in the lower panel of \ref{fig:transport_estimate}. The results here represent a a good qualitative fit and overall similar values to the same heat flux terms evaluated from experimental data. The overall good agreement between simulation and experiment of the four terms modeled in EMC3-EIRENE supports the conclusion that the majority of observed differences between experimental and modeled profiles in this intermediate density range can be explained by a missing poloidal transport term in the code. We conclude based on the interpretation of the experimental data that this terms is a result of $\textbf{E} \times \textbf{B}$ drifts. While the strongest term in the EMC3-EIRENE simulation is the parallel convective term, that is not to say that these calculations imply hollow profiles. As mentioned before, this work does not make predictive assumptions about the absolute ratio of $q_\perp / q_\parallel$ which leads to profile hollowness. Rather, this work seeks to explore the scaling of perpendicular and parallel transports, and provide evidence that two distinct transport regimes are plausible. In this EMC3-EIRENE scan, the $T_e$ profiles are barely hollow at $\overline{n}_e = 0.7 \times 10^{19}$ m$^{-3}$, and quickly become purely monotonic in nature. This corresponds to relatively constant parallel transport (the sum of the parallel conductive and convective traces) and increasing radial transport, leading to a progressive increase in the ratio $q_\perp / q_\parallel$.

The apparent density at which the $\textbf{E} \times \textbf{B}$ drift term becomes sub-dominant, $\overline{n}_e \approx 5.5 \times  10^{19}$ m$^{-3}$, should not be interpreted  as a quantitative prediction of when the hollow $T_e$ profiles vanish. Instead, the experimental measurements provide direct evidence for a strong change in the ratio between perpendicular diffusive and poloidal $\textbf{E} \times \textbf{B}$ drift transport as a function of line-averaged density. This is consistent with the observation of a shift from hollow to monotonic $T_e$ profiles: at some point in the density scan, the increasing edge density gradients enable perpendicular diffusion to transfer energy into the O-point region of the island more efficiently than the poloidal $\textbf{E} \times \textbf{B}$ drifts are capable of convecting heat around the island flux surfaces poloidally. In summary, this analysis confirms that $\textbf{E} \times \textbf{B}$ drifts induced by the observed hollow temperature profiles would be strong enough to maintain hollow temperature profiles in the island SOL, but only up to a certain level of radial diffusive heat flux. 

Lastly, it should be noted that there is an inherent contradiction between the $T_e \approx T_{\text{target}}$ assumption used in calculating the radial electric fields and the presence of a finite conduction term, which explicitly requires non-zero parallel temperature gradients. However, the experimentally observed parallel electron temperature gradients are on the order of $0.1$ eV/m, and as such the total temperature drop into the pre-sheath is generally small. Therefore, the overall impact on the $\textbf{E}_r$ profiles is also small. We also note the small level of parallel conduction term above $\overline{n}_e \approx 2 \times 10^{19}$ m$^{-3}$ in figure \ref{fig:transport_estimate}, which implies that parallel temperature gradients are small compared to the radial temperature gradient which sets the radial electric field. Possible future work to compare to this interpretation of these experimental results can include direct measurements of the plasma potential with Langmuir probes. This could further the understanding of the electrostatic potentials in the island and the applicability of a sheath potential model as used in this work.

\section{Conclusions}

Within this work, the presence of a radial electric field $\textbf{E}_r$ in the island, as derived from a sheath boundary assumption along the entire flux tube, has been shown to be of sufficient strength to be the dominant transport term poloidally around the island at low density. However, at high density, these same heat flux calculations show that a large perpendicular diffusive heat transport channel is capable of washing out the radial $T_e$ gradient required to sustain the drift transport channel. This transition is consistent with the approximate density range in which the hollow profiles are observed to vanish. We have shown experimental evidence for the existence of a $\textbf{E} \times \textbf{B}$ drift-dominated local transport regime in the island divertor of Wendelstein 7-X. This results in hollow $T_e$ profiles as heat is preferentially convected around the island poloidally rather than transported through the island cross-field by anomalous diffusion. 

Experimental data in the form of 2D radial-poloidal $T_e(R,Z)$ profiles show systematic hollowness up to a threshold density of approximately $\overline{n}_e \approx 5 \times 10^{19}$ m$^{-3}$, above which a transition to monotonically decreasing $T_e$ profiles is observed. The specific shift in dominance from the drift term to the perpendicular term at some moderate density is qualitatively supported by an EMC3-EIRENE analysis, which showed hollow profiles are washed out by strong perpendicular transport. In this analysis, significant qualitative disagreements to the experimental measurements in the $T_e$ profiles are seen at low and moderate densities, and the differences were ascribed to electrostatic drifts neglected by the code. $T_e$ profile hollowness was achieved in the simulations only by artificially increasing parallel transport by going to very low SOL density and very high SOL temperature. This is consistent with the predictions of a simple analytical two-point model for stellarators \cite{feng2011comparison}. EMC3-EIRENE simulations were shown to be consistent with experiment only at higher densities, at which the poloidal $\textbf{E} \times \textbf{B}$ transport term becomes small compared to radial anomalous diffusion.

Experimental evidence of significant $E\times B$ drifts has been documented in the past on W7-X, focusing on the ``low iota'' magnetic configuration of the device \cite{kriete2023effects, hammond2019drift}. That configuration is unique for its very low $\Theta$ values and a correspondingly large impact of the poloidal drift transport. This study of the transition between hollow and monotonic temperature profiles provides evidence that SOL transport behavior is strongly influenced by drifts also in the ``standard'' magnetic configuration, at least up to moderate density. As this is the most commonly used magnetic configuration of the device, understanding $\textbf{E} \times \textbf{B}$ is then important for understanding the overall divertor performance of the device, and further study is motivated. 

Strong convection in a reactor-relevant regime may be undesirable, as the convected fluxes may be in a direction away from the pumping fixtures of the divertor, hampering particle exhaust. As such, further experimental investigations into density and power dependencies of this poloidal transport channel are crucial, and are a high priority for the next operational phase of the device and for simulation of future divertor solutions for other devices. Near-future code expansions of EMC3-EIRENE to include ad-hoc convective terms, if not a complete self-consistent modeling of $\textbf{E} \times \textbf{B}$ drifts, offer a promising first step in confirming that these experimental measurements are indeed explainable by poloidal drift effects of similar intensity to the inferred $v_{\textbf{E} \times \textbf{B}} \approx 3$ km/s poloidally around the island. 

Lastly, the findings discussed here are specific for this set of experiments and do not indicate that $\textbf{E} \times \textbf{B}$ drifts are generally negligible in future high density, high heating power scenarios. The precise density at which the transition between hollow and monotonic radial profiles occurs is dependent on the energy flux into the SOL, which is small for these relatively low heating power plasmas, and can be increased in the future. With higher heating power, it is likely that the dominance of the $\textbf{E} \times \textbf{B}$ drift transport will be maintained at higher density, and that this term may also be relevant in these scenarios when detachment is obtained.  As such, a key expansion of this study would include large heating power scans at fixed density, to probe the impact of changing upstream parameters on profile shape.

\section*{Acknowledgements}

This work has been carried out within the framework of the EUROfusion Consortium, funded by the European Union via the Euratom Research and Training Programme (Grant Agreement No 101052200 — EUROfusion). Views and opinions expressed are however those of the author(s) only and do not necessarily reflect those of the European Union or the European Commission. Neither the European Union nor the European Commission can be held responsible for them. This work was funded in part by the U.S. Department of Energy under grant numbers DE-SC00014210 and DE-SC00013911. 

\printbibliography

\end{document}